\documentclass[aps,prb,twocolumn,showpacs,floatfix]{revtex4}

\usepackage{graphicx}

\begin{document}

\title{Excitations near the boundary between a metal and a Mott insulator}

\author{A. Sherman}
\author{N. Voropajeva}
 \affiliation{Institute of Physics, University of Tartu, Riia 142,
 51014 Tartu, Estonia}

\date{\today}

\begin{abstract}
A heterostructure of a semi-infinite metal and a Mott insulator is considered. It is supposed that both materials have an identical lattice spacing and hopping integrals and differ in the Hubbard repulsion which is negligible in the metal and exceeds the critical value for the Mott transition in the insulator. At half-filling and for low temperatures the insulator has the long-range antiferromagnetic order. Its low-lying elementary excitations are standing spin waves and a spin-wave mode which is localized near the interface and has a two-dimensional dispersion. This mode ejects bulk modes from the boundary region. The antiferromagnetic ordering of the insulator induces an antiferromagnetic order in the metal where the magnetization decays exponentially with distance from the interface. This decay is characterized by the correlation length equal to 5--6 lattice spacings.
\end{abstract}


\maketitle

\section{Introduction}
In the last few years an active interest is taken in heterostructures fabricated out of strongly correlated systems. Looking for new effects and their possible applications a wide variety of systems has been investigated both experimentally and theoretically. In particular, it was established that the interface of Mott and band insulators can demonstrate metallic behavior. \cite{Ohtomo02,Ohtomo04,Okamoto,Kancharla} The similar behavior is expected for the interface of two Mott insulators.\cite{Lee} With lowering temperature this two-dimensional (2D) metal becomes superconducting and the transition has some properties of the Berezinsky-Kosterlitz-Thouless transition.\cite{Reyren,Gariglio} The conduction layer can be manipulated by gate voltages which thereby provides a way for carrier doping by electric field, in a disorder-free way, and is promising for new devices.\cite{Nakamura} Notice also experiments with ultracold atoms where the coexistence of several phases with phase boundaries is often observed.\cite{Bloch,Snoek}

In this paper, we study elementary excitations near the boundary of a metal and a Mott insulator which have identical simple cubic lattices. The boundary is perpendicular to one of the crystallographic axes. It is supposed that the two crystals have identical hopping integrals and differ only in the value of the Hubbard repulsion which is vanishingly small in the metal and exceeds the critical value $U_c\approx 2.8B$ in the Mott insulator. Here $B=6|t|$ is the halfwidth of the electron band in the considered case when only the hopping integral $t$ between nearest neighbor sites is nonzero. At half-filling the critical value $U_c$ separates the metallic and the insulating phases in the bulk.\cite{Georges} In these conditions, the penetration of the metallic state into the Mott insulator is negligibly small.\cite{Helmes} When the temperature is lower than the Ne\'el temperature the semi-infinite insulator has the long-range antiferromagnetic order. We investigate its spin excitations which are the only low-lying excitations in the considered case. The small ratio $|t|/U$ allows us to simplify the initial problem, reducing it to the interface between a metal and a Heisenberg antiferromagnet. The spectrum of the antiferromagnet consists of standing spin waves with the dispersion of the bulk three-dimensional (3D) spin waves and a mode of surface spin waves which is localized within a few layers near the surface and has a 2D dispersion. The bandwidth of the surface mode is larger than that in the pure 2D case with the same exchange constant, though it is smaller than the bandwidth in the 3D case. The mathematical description of the surface mode has much in common with the description of point defect states and, as in this latter problem, the localized states (the surface mode) eject the bulk states (standing waves) from the region near the surface. Then we consider the influence of the antiferromagnetic ordering of the Mott insulator on the magnetic state of the metal. We found that the region of the metal near the boundary is also antiferromagnetically ordered. The magnetization in the metal decays exponentially with distance from the boundary. The correlation length is equal to 5--6 lattice spacings and is nearly independent of the ratio $|t|/U$.

The description of the considered model and the derivation of the effective Hamiltonian are given in Sec.~II. In Sec.~III, the eigenvalues and eigenvectors of Hamiltonians of the semi-infinite metal and Heisenberg antiferromagnet, which are two parts of the effective Hamiltonian, are considered. In Sec.~IV, the magnetic ordering in the metal, which is caused by the antiferromagnetic ordering of the insulator, is studied. Concluding remarks are presented in Sec.~V.

\section{Model}
The axes are chosen in such a way that the metal is located in the half-space $l_x\leq -1$, while the Mott insulator is in the half-space $l_x\geq 0$. Here $l_x$, $l_y$ and $l_z$ label sites of a 3D cubic lattice and the lattice spacing is set as the unit of length. For the half-filled case the Hamiltonian of the system reads
\begin{eqnarray}\label{Hamiltonian}
H&=&t\sum_{<\bf LL'>}\sum_\sigma a^\dagger_{\bf L\sigma}a_{\bf L'\sigma}\nonumber\\
&+&U\sum_{{\bf l},l_x\geq 0}\left(n_{\bf L\uparrow}-\frac{1}{2}\right) \left(n_{\bf L\downarrow}-\frac{1}{2}\right),
\end{eqnarray}
where ${\bf L}=(l_x,l_y,l_z)$, ${\bf l}=(l_y,l_z)$, the notation $<\bf LL'>$ in the first sum indicates the summation over all pairs of nearest neighbor sites, $a^\dagger_{\bf L\sigma}$ is the electron creation operator with the spin projection $\sigma=\uparrow$ or $\downarrow$, $n_{\bf L\sigma}=a^\dagger_{\bf L\sigma}a_{\bf L\sigma}$, and $U$ is the Hubbard repulsion which, as mentioned above, is supposed to be larger than $U_c$ and much larger than $|t|$.

In these conditions the Mott region of the system is an insulator with the spin excitations being the only low-lying excitations. This fact allows us to simplify essentially the further consideration by using a unitary transformation which eliminates terms of the first order in $t$ that change the occupation of the Hubbard subbands in the Mott region. This unitary transformation is similar to that used for the derivation of the $t$-$J$ Hamiltonian from the Hubbard Hamiltonian.\cite{Hirsch} For this purpose it is convenient to switch to the Hubbard operators\cite{Izyumov,Ovchinnikov} $X^{ij}_{\bf L}=|{\bf L}i\rangle\langle{\bf L}j|$ in the Mott region. Here $|{\bf L}i\rangle$ denotes four possible site states -- the unoccupied state ($i=0$), the two singly occupied states ($i=\sigma$), and the doubly occupied state ($i=2$). The electron creation and annihilation operators are connected with the Hubbard operators by the relations
\begin{equation}\label{Xoperators}
a_{\bf L\sigma}=X^{0\sigma}_{\bf L}+\sigma X^{-\sigma,2}_{\bf L},\quad a^\dagger_{\bf L\sigma}=X^{\sigma 0}_{\bf L}+\sigma X^{2,-\sigma}_{\bf L}.
\end{equation}
Here we use an alternative definition of $\sigma$: $\sigma=\pm 1$. In these notations Hamiltonian (\ref{Hamiltonian}) reads
\begin{eqnarray}
H&=&H_0+H_1+H_2,\quad H_0=\frac{U}{2}\sum_{{\bf l},l_x\geq 0}\left( X^{00}_{\bf L}+X^{22}_{\bf L}\right),\nonumber\\
H_1&=&t\sum_{\bf la\sigma}\sum_{l_x\leq -1}a^\dagger_{{\bf l+a},l_x \sigma}a_{{\bf l}l_x\sigma}\nonumber\\
&+&t\sum_{\bf l\sigma}\sum_{l_x\leq -1}\left(a^\dagger_{{\bf l},l_x-1, \sigma}a_{{\bf l}l_x\sigma}+a^\dagger_{{\bf l},l_x \sigma}a_{{\bf l},l_x-1,\sigma}\right)\nonumber\\
&+&t\sum_{\bf la\sigma}\sum_{l_x\geq 0}\left(X^{\sigma 0}_{{\bf l+a},l_x} X^{0\sigma}_{{\bf l}l_x}+X^{2,-\sigma}_{{\bf l+a},l_x} X^{-\sigma,2}_{{\bf l}l_x}\right)\nonumber\\
&+&t\sum_{\bf l\sigma}\sum_{l_x\geq -1}\left(X^{\sigma 0}_{{\bf l},l_x+1} X^{0\sigma}_{{\bf l}l_x}+X^{2,-\sigma}_{{\bf l},l_x+1} X^{-\sigma,2}_{{\bf l}l_x}\right.\label{Hamiltonian2}\\
&&\quad\quad+\left. X^{\sigma 0}_{{\bf l}l_x} X^{0\sigma}_{{\bf l},l_x+1}+X^{2,-\sigma}_{{\bf l}l_x} X^{-\sigma,2}_{{\bf l},l_x+1}\right),\nonumber\\
H_2&=&t\sum_{\bf la\sigma}\sum_{l_x\geq 0}\sigma\left(X^{\sigma 0}_{{\bf l+a},l_x} X^{-\sigma,2}_{{\bf l}l_x}+X^{2,-\sigma}_{{\bf l+a},l_x} X^{0\sigma}_{{\bf l}l_x}\right)\nonumber\\
&+&t\sum_{\bf l\sigma}\sum_{l_x\geq -1}\sigma\left(X^{\sigma 0}_{{\bf l},l_x+1} X^{-\sigma,2}_{{\bf l}l_x}+X^{2,-\sigma}_{{\bf l},l_x+1} X^{0\sigma}_{{\bf l}l_x}\right.\nonumber\\
&&\quad\quad+\left. X^{\sigma 0}_{{\bf l}l_x} X^{-\sigma,2}_{{\bf l},l_x+1}+X^{2,-\sigma}_{{\bf l}l_x} X^{0\sigma}_{{\bf l},l_x+1}\right),\nonumber
\end{eqnarray}
where ${\bf a}=(\pm 1,0),(0,\pm 1)$ are four unitary vectors connecting nearest neighbor sites in the YZ plane.

The unitary transformation we are looking for has to remove terms of the first order in $t$ which change the occupation of the Hubbard subbands in the Mott region from the transformed Hamiltonian. Terms of this type are collected in the part $H_2$ in Eq.~(\ref{Hamiltonian2}). Up to the terms of the second order in $t$ the transformed Hamiltonian can be written as
\begin{eqnarray}\label{transformation}
\tilde{H}&=&e^SHe^{-S}\nonumber\\
&\approx&H_0+H_1+H_2+[S,H_0]+[S,H_1]+[S,H_2]\nonumber\\
&&+\frac{1}{2}[S,[S,H_0]].
\end{eqnarray}
The operator $S$ is looked for in the form
\begin{eqnarray}
S&=&\xi\sum_{\bf la\sigma}\sum_{l_x\geq 0}\sigma\left(X^{\sigma 0}_{{\bf l+a},l_x} X^{-\sigma,2}_{{\bf l}l_x}-X^{2,-\sigma}_{{\bf l+a},l_x} X^{0\sigma}_{{\bf l}l_x}\right)\nonumber\\
&+&\xi\sum_{\bf l\sigma}\sum_{l_x\geq 0}\sigma\left(X^{\sigma 0}_{{\bf l},l_x+1} X^{-\sigma,2}_{{\bf l}l_x}-X^{2,-\sigma}_{{\bf l},l_x+1} X^{0\sigma}_{{\bf l}l_x}\right.\nonumber\\
&&\quad\quad+\left. X^{\sigma 0}_{{\bf l}l_x} X^{-\sigma,2}_{{\bf l},l_x+1}-X^{2,-\sigma}_{{\bf l}l_x} X^{0\sigma}_{{\bf l},l_x+1}\right)\nonumber\\
&+&\xi'\sum_{\bf l\sigma}\sigma\left(X^{\sigma 0}_{{\bf l}0} X^{-\sigma,2}_{{\bf l},-1}-X^{2,-\sigma}_{{\bf l}0} X^{0\sigma}_{{\bf l},-1}\right.\nonumber\\
&&\quad\quad+\left. X^{\sigma 0}_{{\bf l},-1} X^{-\sigma,2}_{{\bf l}0}-X^{2,-\sigma}_{{\bf l},-1} X^{0\sigma}_{{\bf l}0}\right),
\label{generator}
\end{eqnarray}
and the parameters $\xi$ and $\xi'$ are determined from the condition
\begin{equation}\label{condition}
H_2+[S,H_0]=0,
\end{equation}
which eliminates the mentioned terms from the transformed Hamiltonian~(\ref{transformation}). Using Eqs.~(\ref{Hamiltonian2}), (\ref{generator}), and (\ref{condition}) we find
\begin{equation}\label{xi}
\xi=-\frac{t}{U},\quad \xi'=-\frac{2t}{U}.
\end{equation}

Finally the transformed Hamiltonian reads
\begin{eqnarray}
\tilde{H}&=&t\sum_{\bf la\sigma}\sum_{l_x\leq -1}a^\dagger_{{\bf l+a},l_x \sigma}a_{{\bf l}l_x\sigma}\nonumber\\
&+&t\sum_{\bf l\sigma}\sum_{l_x\leq -1}\left(a^\dagger_{{\bf l},l_x-1, \sigma}a_{{\bf l}l_x\sigma}+a^\dagger_{{\bf l},l_x \sigma}a_{{\bf l},l_x-1,\sigma}\right)\nonumber\\
&+&\frac{J}{2}\sum_{\bf la}\sum_{l_x\geq 0}{\bf S}_{{\bf l+a},l_x}{\bf S}_{{\bf l}l_x}+J\sum_{{\bf l},l_x\geq 0}{\bf S}_{{\bf l},l_x+1}{\bf S}_{{\bf l}l_x}\nonumber\\
&+&2J\sum_{\bf l}\left({\bf S}_{{\bf l}0}{\bf S}_{{\bf l},-1}-\frac{1}{4}\sum_\sigma X^{\sigma\sigma}_{{\bf l},-1}\right), \label{Hamiltonian3}
\end{eqnarray}
where $J=4t^2/U$, the components of the spin-$\frac{1}{2}$ vector ${\bf S_L}$ are $S^\sigma_{\bf L}=X^{\sigma,-\sigma}_{\bf L}$ and $S^z_{\bf L}=\frac{1}{2}(X^{\uparrow\uparrow}_{\bf L}-X^{\downarrow\downarrow}_{\bf L})$. In Hamiltonian (\ref{Hamiltonian3}), we have neglected terms of the second order in $t$, which change the occupation of the Hubbard subbands in the Mott region or give a correction to the kinetic energy in the boundary layer $l_x=-1$. We neglected also terms which describe the intrasubband transport in the Mott region. At half-filling such processes are suppressed.

Notice that the spin bond between the boundary layers $l_x=-1$ and $l_x=0$ is twice as much the bond in the Mott region [in the third term of Hamiltonian (\ref{Hamiltonian3}) each bond appears twice in the sum].

\section{Semi-infinite metal and antiferromagnet}
Let us first consider eigenstates of a semi-infinite metal and a Heisenberg antiferromagnet which Hamiltonians are contained in Eq.~(\ref{Hamiltonian3}). The Hamiltonian of the metal reads
\begin{eqnarray}\label{Hmetal}
H_m&=&t\sum_{\bf k\sigma}\sum_{l_x\leq -1}\left(4\gamma^{(2)}_{\bf k}a^\dagger_{{\bf k}l_x\sigma}a_{{\bf k}l_x\sigma}\right.\nonumber\\
&&\quad\quad\left.+a^\dagger_{{\bf k},l_x-1,\sigma}a_{{\bf k}l_x\sigma}+a^\dagger_{{\bf k}l_x\sigma}a_{{\bf k},l_x-1,\sigma}\right),
\end{eqnarray}
where the translational invariance of the system in the YZ plane was taken into account,
$$a_{{\bf k}l_x\sigma}=\frac{1}{\sqrt{N}}\sum_{\bf l}e^{-i{\bf lk}}a_{{\bf l}l_x\sigma},$$ $N$ is the number of sites in the periodic YZ region, {\bf k} is the 2D wave vector, and $\gamma^{(2)}_{\bf k}=\frac{1}{4}\sum_{\bf a}e^{i{\bf ka}}$. Hamiltonian (\ref{Hmetal}) is diagonalized by the unitary transformation
\begin{equation}\label{transf1}
a_{{\bf k}l_x\sigma}=\sum_{k_x}\alpha_{l_xk_x}a_{{\bf k}k_x\sigma},
\end{equation}
where $\alpha_{l_xk_x}$ satisfies the conditions
\begin{equation}\label{uconditions}
\sum_{k_x}\alpha_{l_xk_x}\alpha^*_{l'_xk_x}=\delta_{l_xl'_x},\quad \sum_{l_x}\alpha_{l_xk_x}\alpha^*_{l_xk'_x}=\delta_{k_xk'_x}.
\end{equation}
Substituting Eq.~(\ref{transf1}) into Hamiltonian (\ref{Hmetal}) and using conditions (\ref{uconditions}) we find the following relation for $\alpha_{l_xk_x}$:
\begin{equation}\label{eqforalpha}
\alpha_{l_x+1,k_x}+\alpha_{l_x-1,k_x}=\left(t^{-1}\varepsilon_{{\bf k}k_x}-4\gamma^{(2)}_{\bf k}\right)\alpha_{l_xk_x}
\end{equation}
with the boundary condition
\begin{equation}\label{boundary}
\alpha_{l_x=0,k_x}=0.
\end{equation}
Here $\varepsilon_{{\bf k}k_x}$ is the eigenvalue of Hamiltonian (\ref{Hmetal}). We seek the solution of Eq.~(\ref{eqforalpha}) in the form $\alpha_{l_xk_x}\sim e^{\kappa(k_x)l_x}$ where $\kappa$ has to be real or purely imaginary for the eigenvalue of Eq.~(\ref{eqforalpha}), $2\cosh[\kappa(k_x)]$, be real. Solutions with real $\kappa$-s do not satisfy condition (\ref{boundary}) and cannot be mixed with solutions with imaginary $\kappa$-s, since they correspond to different energies. Only a linear combination of two solutions corresponding to the same eigenenergy (for a fixed {\bf k}) form a new solution. The linear combination of two solutions with imaginary and opposite in sign $\kappa$-s, the standing wave, satisfies condition (\ref{boundary}),
\begin{eqnarray}
\alpha_{l_xk_x}&=&\sqrt{\frac{2}{\pi}}\sin(k_xl_x),\nonumber\\[-0.2em]
&&\label{solution}\\[-0.2em]
\varepsilon_{{\bf k}k_x}&=&2[\cos(k_x)+\cos(k_y)+\cos(k_z)],\nonumber
\end{eqnarray}
where $k_x$ varies continuously in the range $(0,\pi)$. Thus, in Eqs.~(\ref{transf1}) and (\ref{uconditions}) sums over $k_x$ have to be substituted with integrals and the Kronecker symbol with the Dirac delta function.

Now let us consider the semi-infinite Heisenberg antiferromagnet described by the Hamiltonian
\begin{equation}\label{Haf}
H_a=\frac{J}{2}\sum_{\bf la}\sum_{l_x\geq 0}{\bf S}_{{\bf l+a},l_x}{\bf S}_{{\bf l}l_x}+J\sum_{{\bf l},l_x\geq 0}{\bf S}_{{\bf l},l_x+1}{\bf S}_{{\bf l}l_x}.
\end{equation}
For low temperatures this system is characterized by the long-range antiferromagnetic ordering. Therefore to describe its low-lying elementary excitations we use the spin wave approximation:
\begin{eqnarray}
S^z_{\bf L}&=&e^{i{\bf\Pi L}}\left(\frac{1}{2}-b^\dagger_{\bf L}b_{\bf L}\right),\nonumber\\[-0.3em]
&&\label{swa}\\[-0.3em]
S^+_{\bf L}&=&P^+_{\bf L}b_{\bf L}+P^-_{\bf L}b^\dagger_{\bf L},\quad S^-_{\bf L}=P^-_{\bf L}b_{\bf L}+P^+_{\bf L}b^\dagger_{\bf L},\nonumber
\end{eqnarray}
where the spin-wave operators $b_{\bf L}$ and $b^\dagger_{\bf L}$ satisfy the Boson commutation relations and
$${\bf\Pi}=(\pi,\pi,\pi),\quad P^\pm_{\bf L}=\frac{1}{2}\left(1\pm e^{i{\bf\Pi L}}\right).$$
Substituting Eq.~(\ref{swa}) into Eq.~(\ref{Haf}) and using the translation invariance of the Hamiltonian in the YZ plane we find
\begin{eqnarray}\label{Haf2}
H_a&=&J\sum_{{\bf k},l_x\geq 0}\left[3\left(1-\frac{1}{6}\delta_{l_x0} \right)b^\dagger_{{\bf k}l_x}b_{{\bf k}l_x}\right.\nonumber\\
&+&\gamma^{(2)}_{\bf k}\left( b_{{\bf k}l_x}b_{{\bf -k},l_x}+b^\dagger_{{\bf k}l_x}b^\dagger_{{\bf -k},l_x}\right)\nonumber\\
&+&\left.\frac{1}{2}\left(b_{{\bf k}l_x}b_{{\bf -k},l_x+1}+ b^\dagger_{{\bf k}l_x}b^\dagger_{{\bf -k},l_x+1}\right)\right],
\end{eqnarray}
where
$$b_{{\bf k}l_x}=\frac{1}{\sqrt{N}}\sum_{\bf l}e^{-i{\bf kl}}b_{{\bf l}l_x}.$$

The usual approach to the diagonalization of biquadratic forms of the type of Eq.~(\ref{Haf2}) is the use of the Bogoliubov-Tyablikov transformation\cite{Tyablikov} which in the present case reads
\begin{equation}\label{BTt}
b_{{\bf k}l_x}=\sum_{k_x}\left(u_{{\bf k}l_xk_x}\beta_{{\bf k}k_x}+v_{{\bf k}l_xk_x}\beta^\dagger_{{\bf -k},k_x}\right),
\end{equation}
where the operators $\beta^\dagger_{{\bf k}k_x}$ and $\beta_{{\bf k}k_x}$ also satisfy the Boson commutation relations and therefore the coefficients $u_{{\bf k}l_xk_x}$ and $v_{{\bf k}l_xk_x}$ satisfy the following conditions:
\begin{eqnarray}
&&\sum_{k_x}\left(u_{{\bf k}l_xk_x}u^*_{{\bf k}l'_xk_x}-v_{{\bf k}l_xk_x}v^*_{{\bf k}l'_xk_x}\right)=\delta_{l_xl'_x},\nonumber\\ &&\sum_{k_x}\left(u_{{\bf k}l_xk_x}v_{{\bf -k},l'_xk_x}-v_{{\bf k}l_xk_x}u_{{\bf -k},l'_xk_x}\right)=0,\nonumber\\[-0.4em]
&&\label{ucond2}\\[-0.4em]
&&\sum_{l_x\geq 0}\left(u_{{\bf k}l_xk_x}u^*_{{\bf k}l_xk'_x}-v_{{\bf -k},l_xk'_x}v^*_{{\bf -k},l_xk_x}\right)=\delta_{k_xk'_x},\nonumber\\ &&\sum_{l_x\geq 0}\left(u^*_{{\bf -k},l_xk_x}v_{{\bf -k}, l_xk'_x}-v_{{\bf k}l_xk_x}u^*_{{\bf k}l_xk'_x}\right)=0.\nonumber
\end{eqnarray}
The transformation which is opposite to Eq.~(\ref{BTt}) reads
\begin{equation}\label{iBTt}
\beta_{{\bf k}k_x}=\sum_{l_x\geq 0}\left(u^*_{{\bf k}l_xk_x}b_{{\bf k}l_x}-v_{{\bf -k},l_xk_x}b^\dagger_{{\bf -k},l_x}\right).
\end{equation}

In the new representation Hamiltonian (\ref{Haf2}) is diagonal,
$$H_a=\sum_{{\bf k}k_x}E_{{\bf k}k_x}\beta^\dagger_{{\bf k}k_x}\beta_{{\bf k}k_x}+{\rm const}.$$
If we use Eqs.~(\ref{Haf2})-(\ref{iBTt}) in the relation
$$[\beta_{{\bf k}k_x},H_a]=E_{{\bf k}k_x}\beta_{{\bf k}k_x},$$
we find equations for the determination of the coefficients $u_{{\bf k}l_xk_x}$, $v_{{\bf k}l_xk_x}$ and the energy $E_{{\bf k}k_x}$,
\begin{eqnarray}
&&E_{{\bf k}k_x}u^*_{{\bf k}l_xk_x}=J\left[3\left(1-\frac{1}{6} \delta_{l_x0}\right)u^*_{{\bf k}l_xk_x}\right.\nonumber\\
&&\quad\quad\left.+2\gamma^{(2)}_{\bf k}v_{{\bf -k},l_xk_x}+\frac{1}{2}\left(v_{{\bf -k},l_x+1,k_x}+v_{{\bf -k},l_x-1,k_x}\right)\right],\nonumber\\[-0.2em]
&&\label{eqforuv}\\[-0.2em]
&&-E_{{\bf k}k_x}v_{{\bf -k},l_xk_x}=J\left[3\left(1-\frac{1}{6} \delta_{l_x0}\right)v_{{\bf -k},l_xk_x}\right.\nonumber\\
&&\quad\quad\left.+2\gamma^{(2)}_{\bf k}u^*_{{\bf k}l_xk_x} +\frac{1}{2}\left(u^*_{{\bf k},l_x+1,k_x}+u^*_{{\bf k},l_x-1,k_x}\right)\right],\nonumber
\end{eqnarray}
with the boundary conditions
\begin{equation}\label{conditions}
u^*_{{\bf k},l_x=-1,k_x}=0,\quad v_{{\bf -k},l_x=-1,k_x}=0.
\end{equation}

If, for the time being, we neglect terms proportional to $\delta_{l_x0}$ in Eq.~(\ref{eqforuv}), solutions for this set of equations can be found in the form
$$u^*_{{\bf k}l_xk_x}\sim e^{\kappa(k_x)l_x},\quad v_{{\bf -k},l_xk_x}\sim e^{\kappa(k_x)l_x},$$
where $\kappa$ has to be either real or purely imaginary for the energy $E_{{\bf k}k_x}$ be real. Again we find that solutions with real $\kappa$-s do not satisfy boundary conditions (\ref{conditions}) and cannot be admixed to solutions with imaginary $\kappa$-s, since these two groups of solutions corresponds to different ranges of energy for a fixed pair $({\bf k},{\bf -k})$. Thus, the solutions of this simplified problem which satisfy boundary conditions (\ref{conditions}) are standing waves
\begin{eqnarray}
&&u^*_{{\bf k}l_xk_x}=A_{{\bf k}k_x}\sin[k_x(l_x+1)],\nonumber\\ &&v_{{\bf -k},l_xk_x}=B_{{\bf k}k_x}\sin[k_x(l_x+1)],\nonumber\\
&&A_{{\bf k}k_x}=\sqrt{\frac{2}{\pi}}\frac{3J+E_{{\bf k}k_x}}{\sqrt{\left(3J+E_{{\bf k}k_x}\right)^2-\left(3J\gamma^{(3)}_{{\bf k}k_x}\right)^2}},\label{swave}\\
&&B_{{\bf k}k_x}=-\sqrt{\frac{2}{\pi}}\frac{3J\gamma^{(3)}_{{\bf k}k_x}}{\sqrt{\left(3J+E_{{\bf k}k_x}\right)^2-\left(3J\gamma^{(3)}_{{\bf k}k_x}\right)^2}},\nonumber\\
&&E_{{\bf k}k_x}=3J\sqrt{1-\left(\gamma^{(3)}_{{\bf k}k_x}\right)^2}, \nonumber
\end{eqnarray}
where $\gamma^{(3)}_{{\bf k}k_x}=\frac{1}{3}[\cos(k_x)+\cos(k_y)+ \cos(k_z)]$ and $k_x$ varies continuously in the range $(0,\pi)$. Thus, in the above formulas, summations over $k_x$ have to be understood as integrations in the indicated limits and the Kronecker symbols of $k_x$ have to be substituted with the Dirac delta functions.

The simple exponential solutions are inapplicable, if we take into account the previously dropped terms which are proportional to $\delta_{l_x0}$. To obtain solutions for this more complicated problem we use a modification of the method applied by I.~M.~Lifshits for the problem of a local defect.\cite{Lifshits} Let us introduce the two-component operator
$$\hat{B}_{{\bf k}l_x}=\left(\begin{array}{c}
b_{{\bf k}l_x}\\
b^\dagger_{{\bf -k},l_x}\end{array}\right)$$
and determine the matrix retarded Green's function
\begin{equation}\label{GfD}
\hat{D}({\bf k}tl_xl'_x)=-i\theta(t)\left\langle\left[\hat{B}_{{\bf k}l_x}(t),\hat{B}^\dagger_{{\bf k}l'_x}\right]\right\rangle,
\end{equation}
where $\hat{B}_{{\bf k}l_x}(t)=e^{iH_at}\hat{B}_{{\bf k}l_x}e^{-iH_at}$ with the $H_a$ determined by Eq.~(\ref{Haf2}). Green's function (\ref{GfD}) satisfies the equation
\begin{eqnarray}
i\frac{d}{dt}\hat{D}({\bf k}tl_xl'_x)&=&\delta(t)\delta_{l_xl'_x} \hat{\tau}_3+ J\biggl[3\left(1-\frac{1}{6}\delta_{l_x0}\right) \hat{\tau}_3\nonumber\\
&+&2\gamma^{(2)}_{\bf k}\hat{\tau}_1\biggr]\hat{D}({\bf k}tl_xl'_x) +\frac{J}{2}\hat{\tau}_1\nonumber\\
&\times&\left[\hat{D}({\bf k}t,l_x-1,l'_x)+\hat{D}({\bf k}t,l_x+1,l'_x)\right],
\nonumber\\
&&\label{eqforD}
\end{eqnarray}
where the matrices $\hat{\tau}_1$ and $\hat{\tau}_3$ are given by
$$\hat{\tau}_1=\left(
           \begin{array}{cc}
             0 & 1 \\
             -1 & 0 \\
           \end{array}
         \right),\quad
\hat{\tau}_3=\left(
         \begin{array}{cc}
           1 & 0 \\
           0 & -1 \\
         \end{array}
       \right).
$$
If we define $\hat{D}^{(0)}({\bf k}tl_xl'_x)$ as Green's function which satisfies Eq.~(\ref{eqforD}) without the term proportional to $\delta_{l_x0}$, the solution of Eq.~(\ref{eqforD}) can be written as
\begin{eqnarray}
&&\hat{D}({\bf k}tl_xl'_x)=\hat{D}^{(0)}({\bf k}tl_xl'_x)\nonumber\\
&&\quad\quad\quad-\frac{J}{2}\int^\infty_{-\infty}dt'\hat{D}^{(0)}({\bf k},t'-t,l_x0)\hat{D}({\bf k}t'0l'_x).\quad\label{eqforD2}
\end{eqnarray}
After the Fourier transformation,
$$\hat{D}({\bf k}\omega l_xl'_x)=\int^\infty_{-\infty}dt e^{i\omega t}\hat{D}({\bf k}tl_xl'_x),$$ we find finally
\begin{eqnarray}
&&\hat{D}({\bf k}\omega l_xl'_x)=\hat{D}^{(0)}({\bf k}\omega l_xl'_x)-\frac{J}{2}\hat{D}^{(0)}({\bf k}\omega l_x0)\nonumber\\
&&\quad\quad\quad\times\left[\hat{\tau}_0+\frac{J}{2}\hat{D}^{(0)}({\bf k}\omega 00)\right]^{-1}\hat{D}^{(0)}({\bf k}\omega 0l'_x),\quad \label{eqforD3}
\end{eqnarray}
where $\hat{\tau}_0$ is the 2$\times$2 unit matrix.

Green's function $\hat{D}^{(0)}({\bf k}\omega l_xl'_x)$ corresponds to the Ha\-mi\-l\-to\-ni\-an which eigenvalues and eigenstates are given by Eq.~(\ref{swave}). This Green's function can be easily calculated,
\begin{eqnarray}
&&\hat{D}^{(0)}({\bf k}\omega l_xl'_x)=\int_0^\pi dk_x\sin[k_x(l_x+1)] \sin[k_x(l'_x+1)]\nonumber\\
&&\quad\quad\times\left(\frac{1}{\omega-E_{{\bf k}k_x}+i\eta} \hat{P}_{{\bf k}k_x}-\frac{1}{\omega+E_{{\bf k}k_x}+i\eta}\hat{Q}_{{\bf k}k_x}\right), \nonumber\\
&&\hat{P}_{{\bf k}k_x}=\left(
 \begin{array}{cc}
 A^2_{{\bf k}k_x} & A_{{\bf k}k_x}B_{{\bf k}k_x} \\
 A_{{\bf k}k_x}B_{{\bf k}k_x} & B^2_{{\bf k}k_x} \\
 \end{array}
\right),\label{eqforD0}\\
&&\hat{Q}_{{\bf k}k_x}=\left(
 \begin{array}{cc}
 B^2_{{\bf k}k_x} & A_{{\bf k}k_x}B_{{\bf k}k_x} \\
 A_{{\bf k}k_x}B_{{\bf k}k_x} & A^2_{{\bf k}k_x} \\
 \end{array}
\right),\nonumber
\end{eqnarray}
where $\eta=+0$.

\begin{figure}[b]
\centerline{\includegraphics[width=7.5cm]{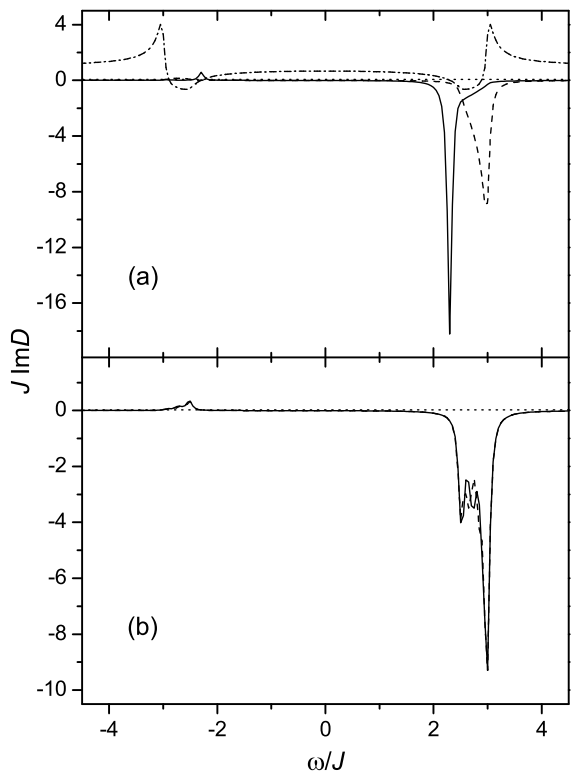}}
\caption{The imaginary parts of Green's functions $D_{11}({\bf k}\omega l_xl_x)$ (the solid lines) and $D^{(0)}_{11}({\bf k}\omega l_xl_x)$ (the dashed lines) for $l_x=0$ (a) and $l_x=5$ (b). ${\bf k}=(0,0.6\pi)$. In part (a), the dash-dotted line demonstrates the real part of the denominator in the second term in the right-hand side of Eq.~(\protect\ref{eqforD3}).} \label{Fig1}
\end{figure}
As follows from Eqs.~(\ref{eqforD3}) and (\ref{eqforD0}), along with the standing waves (\ref{swave}) Green's function $\hat{D}({\bf k}\omega l_xl'_x)$ may have poles corresponding to the surface spin waves. These poles are connected with the second term in the right-hand side of Eq.~(\ref{eqforD3}) and they are perceptible in the spectral function only near the surface of the antiferromagnet, since Green's functions $\hat{D}^{(0)}({\bf k}\omega l_x0)$ and $\hat{D}^{(0)}({\bf k}\omega 0l'_x)$ in this term decrease rapidly with increasing the distance $l_x$, $l'_x$ from the surface. In Fig.~\ref{Fig1}, the spectral function of spin excitations ${\rm Im}D_{11}({\bf k}\omega l_xl_x)$ is shown in comparison with the spectral function of the standing waves ${\rm Im}D^{(0)}_{11}({\bf k}\omega l_xl_x)$ for different distances from the surface. Directly on the surface a pronounced peak at $\omega_0\approx 2.3J$ and an analogous but less intensive (for the given {\bf k}) peak at $-\omega_0$ are observed in the spectrum. From Fig.~\ref{Fig1}(a) it is seen that the peaks are manifestations of the poles of the second term in the right-hand side of Eq.~(\ref{eqforD3}) and therefore correspond to the surface mode. Already for $l_x=5$ the peaks become indistinguishable in the spectrum and the spectra ${\rm Im}D_{11}({\bf k}\omega l_xl_x)$ and ${\rm Im}D^{(0)}_{11}({\bf k}\omega l_xl_x)$ become practically identical [see Fig.~\ref{Fig1}(b)].
\begin{figure}
\centerline{\includegraphics[width=7.5cm]{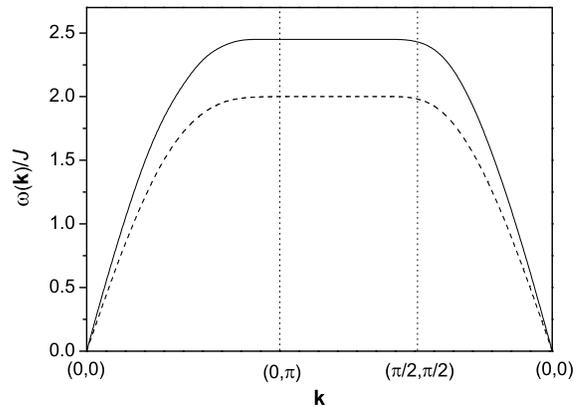}}
\caption{The dispersion of the surface spin-wave mode along the symmetry lines of the Brillouin zone (the solid line). The dashed line demonstrates the dispersion of the two-dimensional spin waves, Eq.~(\protect\ref{2dsw}).} \label{Fig2}
\end{figure}

The similarity in the description of impurity states and the surface mode leads to some similarity in their properties. As seen from Fig.~\ref{Fig1}(a), nearly all quasiparticle weight of the spectrum is concentrated in the peak of the surface mode, while bulk modes manifest themselves as a weak shoulder to its right -- as in the impurity problem,\cite{Lifshits} localized states (the surface mode) eject bulk states (standing waves) from the region near the surface.

Figure~\ref{Fig2} demonstrates the dispersion of the surface spin waves derived from the spectral function ${\rm Im}D_{11}({\bf k}\omega l_xl_x)$ for $l_x=0$. The dispersion (the solid line) is shown along the symmetry lines of the Brillouin zone. The shape of the curve resembles the dispersion of the 2D spin waves,\cite{Izyumov}
\begin{equation}\label{2dsw}
\omega_{\bf k}=2J\sqrt{1-\left(\gamma^{(2)}_{\bf k}\right)^2},
\end{equation}
shown with the dashed line in the figure. However, the frequencies of the surface excitations are somewhat higher, though their bandwidth is smaller than the bandwidth $3J$ of the 3D spin waves [see  Eq.~(\ref{swave})].

In the zeroth order values of the spin operators ${\bf S}_{{\bf l},-1}$ on the metal side of the considered heterostructure vanish. Therefore in this approximation we can neglect the influence of the last term in the Hamiltonian~(\ref{Hamiltonian3}) on the spin-waves of the antiferromagnet and consider the obtained solutions as excitations of the antiferromagnetic part of the heterostructure.

\section{Magnetic ordering in the metal}
Now let us consider the influence of the antiferromagnetic ordering of the Mott insulator on the magnetic state of the metal. In accord with the idea of the spin-wave approximation, which considers the spin-wave operators $b_{\bf L}$ and $b^\dagger_{\bf L}$ as describing small deviations from an equilibrium orientation, we can keep the largest terms contained in the longitudinal part of the interaction,
\begin{equation}\label{Hli}
H_{li}=2J\sum_{\bf l}S^z_{{\bf l}0}S^z_{{\bf l},-1}
\end{equation}
in the last row of Eq.~(\ref{Hamiltonian3}). This sum contains the zero-order terms of the spin-wave approximation $S^z_{{\bf l}0}=\frac{1}{2}e^{i{\bf Ql}}$ where ${\bf Q}=(\pi,\pi)$ [see Eq.~(\ref{swa})]. Thus, we shall consider the Hamiltonian
\begin{eqnarray}\label{Hma}
H_{ma}&=&t\sum_{\bf la\sigma}\sum_{l_x\leq -1}a^\dagger_{{\bf l+a},l_x\sigma}a_{{\bf l}l_x\sigma}\nonumber\\
&+&t\sum_{\bf l\sigma}\sum_{l_x\leq -1}\left(a^\dagger_{{\bf l},l_x-1,\sigma}a_{{\bf l}l_x\sigma}+a^\dagger_{{\bf l}l_x\sigma}a_{{\bf l},l_x-1,\sigma}\right)\nonumber\\
&+&\frac{J}{2}\sum_{\bf l\sigma}e^{i{\bf Ql}}\sigma a^\dagger_{{\bf l},-1,\sigma}a_{{\bf l},-1,\sigma},
\end{eqnarray}
where the relation $S^z_{{\bf l},-1}=\frac{1}{2}\sum_\sigma \sigma a^\dagger_{{\bf l},-1,\sigma}a_{{\bf l},-1,\sigma}$ was used.

The parameter we wish to consider is the magnetization in the metal region,
\begin{equation}\label{magnetization}
M_{\bf L}=n_{\bf L\uparrow}-n_{\bf L\downarrow}.
\end{equation}
It is convenient to divide the YZ plane into two sublattices which contain sites with odd and even sums $l_y+l_z$ of the lattice coordinates. The Fourier transformation of the electron annihilation operators over the sites of one of these sublattices reads
$$a_{{\bf q}ml_x\sigma}=\sqrt{\frac{2}{N}}\sum_{{\bf l}(m)}e^{-i{\bf ql}(m)}a_{{\bf l}(m)l_x\sigma},\quad m=1,2,$$
where ${\bf l}(1)$ and ${\bf l}(2)$ label sites of the two sublattices, and the wave vector {\bf q} belongs to the first magnetic Brillouin zone which is half as much as the usual 2D Brillouin zone. These annihilation operators are connected with the Fourier-transformed annihilation operators used above by the relations
\begin{eqnarray}
a_{{\bf q}l_x\sigma}&=&\frac{1}{\sqrt{2}}\left(a_{{\bf q}2l_x\sigma}+ a_{{\bf q}1l_x\sigma}\right),\nonumber\\[-0.4em]
&&\label{eoperators}\\[-0.4em]
a_{{\bf q'}l_x\sigma}&=&\frac{1}{\sqrt{2}}\left(a_{{\bf q}2l_x\sigma}- a_{{\bf q}1l_x\sigma}\right),\nonumber
\end{eqnarray}
where ${\bf q'}$ is a wave vector in the second magnetic Brillouin zone which is connected with the {\bf q} by the relation ${\bf q'}={\bf q}+(\pm\pi,\pm\pi)$.

Let us determine the matrix retarded Green's functions
\begin{equation}\label{GfG}
G_{mm'}({\bf q}tl_xl'_x\sigma)=-i\theta(t)\left\langle\left\{a_{{\bf q}ml_x\sigma}(t),a^\dagger_{{\bf q}m'l'_x\sigma}\right\}\right\rangle
\end{equation}
with the operator time dependence determined by Hamiltonian (\ref{Hma}). The magnetization (\ref{magnetization}) is connected with the Fourier transform of this Green's function by the relation
\begin{eqnarray}\label{magnetization2}
&&M_{{\bf l}(m)l_x}=\nonumber\\
&&\quad\frac{2}{N}\sum_{\bf q}\int_{-\infty}^\infty\!\!\frac{d\omega}{\pi} \frac{{\rm Im}\left[G_{mm}({\bf q}\omega l_xl_x\!\!\downarrow)- G_{mm}({\bf q}tl_xl_x\!\!\uparrow)\right]}{e^{\beta\omega}+1},\nonumber\\
\end{eqnarray}
where $\beta$ is the inverse temperature and the summation over {\bf q} is over the first magnetic Brillouin zone.

\begin{figure}
\centerline{\includegraphics[width=7.5cm]{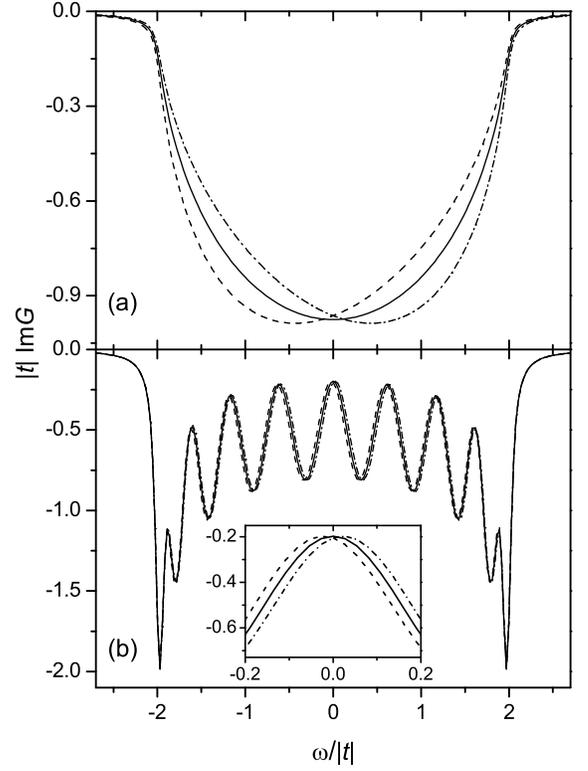}}
\caption{The imaginary parts of Green's functions $G^{(0)}_{11}({\bf q}l_xl_x)$ (the solid lines) and $G_{11}({\bf q}l_xl_x\sigma)$ with $\sigma=\uparrow$ (the dashed lines) and $\sigma=\downarrow$ (the dash-dotted lines) for ${\bf q}=(0,\pi)$, $J=0.23|t|$, the temperature $T=0$, $l_x=-1$ (a), and $l_x=-10$ (b). The inset in part (b) demonstrates the same curves on an enlarged scale.} \label{Fig3}
\end{figure}
\begin{figure}
\centerline{\includegraphics[width=7.5cm]{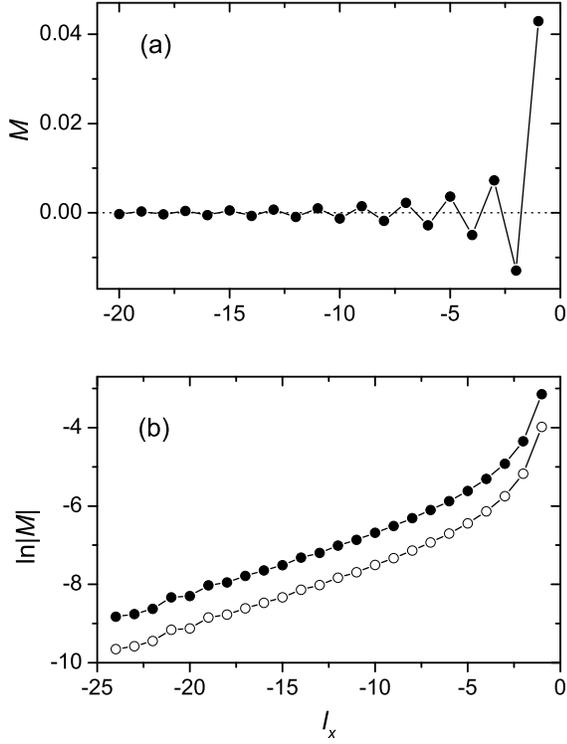}}
\caption{(a) The magnetization $M$ in the metal as a function of the distance from the surface $l_x$ for $J=0.23|t|$ and $T=0$. (b) The absolute value of $M$ in a semi-logarithmic scale for $T=0$, $J=0.23|t|$ (filled circles) and $J=0.1|t|$ (open circles).} \label{Fig4}
\end{figure}
The equation for Green's function (\ref{GfG}) reads
\begin{eqnarray}
i\frac{d}{dt}\hat{G}({\bf q}tl_xl'_x\sigma)&=&\delta(t)\delta_{l_xl'_x}\hat{\tau}_0 \nonumber\\
&+&\left[4t\gamma^{(2)}_{\bf q}\hat{\tau}_2-\frac{J}{2}\sigma \delta_{l_x,-1}\hat{\tau}_3\right]\hat{G}({\bf q}tl_xl'_x\sigma)\nonumber\\
&+&t\Bigl[\hat{G}({\bf q}t,l_x-1,l'_x\sigma)\nonumber\\
&&\quad+\hat{G}({\bf q}t,l_x+1,l'_x\sigma)\Bigr],\label{eqforG}
\end{eqnarray}
where
$$\hat{\tau}_2=\left(
    \begin{array}{cc}
      0 & 1 \\
      1 & 0 \\
    \end{array}
  \right).
$$
Equation (\ref{eqforG}) can be rewritten as
\begin{eqnarray}
&&\hat{G}({\bf q}tl_xl'_x\sigma)=\hat{G}^{(0)}({\bf q}tl_xl'_x)\nonumber\\
&&\quad\quad-\frac{J}{2}\sigma\int_{-\infty}^\infty dt'\hat{G}^{(0)}({\bf q},t-t',l_x,-1)\hat{\tau}_3\nonumber\\
&&\quad\quad\times\hat{G}({\bf q}t',-1,l'_x\sigma),\label{eqforG2}
\end{eqnarray}
and finally for the Fourier transform of Green's function we find
\begin{eqnarray}\label{eqforG3}
&&\hat{G}({\bf q}\omega l_xl'_x\sigma)=\hat{G}^{(0)}({\bf q}\omega l_xl'_x)-\frac{J}{2}\sigma\hat{G}^{(0)}({\bf q}\omega l_x,-1)\nonumber\\
&&\quad\quad\times\hat{\tau}_3\left[\hat{\tau}_0+\frac{J}{2}\sigma \hat{G}^{(0)}({\bf q}\omega,-1,-1)\hat{\tau}_3\right]^{-1}\nonumber\\
&&\quad\quad\times\hat{G}^{(0)}({\bf q}\omega,-1,l'_x).
\end{eqnarray}
In Eqs.~(\ref{eqforG2}) and (\ref{eqforG3}), $\hat{G}^{(0)}({\bf q}\omega l_xl'_x)$ is the Green's function (\ref{GfG}) in which the operator time dependence and the averaging are determined by Hamiltonian (\ref{Hmetal}). It is identical to Hamiltonian (\ref{Hma}) without the last term in the right-hand side. Using Eqs.~(\ref{solution}) and (\ref{eoperators}) we find
\begin{eqnarray}
&&G^{(0)}_{mm'}({\bf q}l_xl'_x)=\int_0^\pi\frac{dk_x}{\pi}\sin(k_xl_x) \sin(k_xl'_x)\nonumber\\
&&\quad\quad\times\left(\frac{1}{\omega-\varepsilon_{{\bf q}k_x}+i\eta}+\frac{(-1)^{m+m'}}{\omega-\varepsilon_{{\bf q'}k_x}+i\eta}\right), \label{eqforG0}
\end{eqnarray}
where ${\bf q'}$, as before, is the wave vector in the second magnetic Brillouin zone, connected with the {\bf q} by the relation ${\bf q'}={\bf q}+(\pm\pi,\pm\pi)$.

Equation~(\ref{eqforG3}) looks similar to Eq.~(\ref{eqforD3}). However, if in the latter equation the multiplier $J$ in the second term of the denominator is of the same order of magnitude as the energy parameter in Green's function $\hat{D}^{(0)}({\bf k}\omega 00)$, in the former equation this multiplier is much smaller than the energy parameter of $\hat{G}^{(0)}({\bf q}\omega,-1,-1)$, the electron band halfwidth $B$, due to the supposition $U\gg|t|$. Therefore surface electronic states do not arise. Nevertheless the influence of the second term in the right-hand side of Eq.~(\ref{eqforG3}) is perceptible, especially near the surface, as seen in Fig.~\ref{Fig3}. It is this term which yields the dependence of Green's function $\hat{G}({\bf q}l_xl'_x\sigma)$ on the spin projection $\sigma$ and leads to a nonzero magnetization (\ref{magnetization2}). As seen from Fig.~\ref{Fig3}(b), for wave vectors near the boundary of the magnetic Brillouin zone this dependence is perceptible even at large distances from the surface which leads to a slow decay of the magnetization with this distance. From Eqs.~(\ref{magnetization2}) and (\ref{eqforG3}) it follows that for a given $l_x$ the magnetization has opposite signs on the two sublattices of the YZ plane.

The dependence of the magnetization on the distance from the surface is shown in Fig.~\ref{Fig4}. We find that the magnetization changes its sign not only on moving from one 2D sublattice to the other but also with transfer perpendicular to the surface -- the magnetization has antiferromagnetic character. As seen from Fig.~\ref{Fig4}, this ordering penetrates into the metal for distance of tens lattice spacings. After the initial rapid decrease the magnetization decays exponentially with $l_x$ with the correlation length $\xi=5-6$ lattice spacings [see Fig.~\ref{Fig4}(b)]. The value of $\xi$ depends only weakly on the ratio $J/|t|$.

Let us check the influence of different corrections to these results. At first let us consider the contribution of the term $b^\dagger_{{\bf l}0}b_{{\bf l}0}$ in the spin component $S^z_{{\bf l}0}$ in the interaction~(\ref{Hli}) [see Eq.~(\ref{swa})]. The value of this contribution can be estimated using the relation
\begin{equation}\label{nb}
\langle b^\dagger_{{\bf l}l_x}b_{{\bf l}l_x}\rangle=\frac{1}{N} \sum_{\bf k}\int^\infty_{-\infty}\frac{d\omega}{\pi}\frac{{\rm Im}D_{11}({\bf k}\omega l_xl_x)}{1-e^{\omega\beta}}.
\end{equation}
Calculations with Green's function obtained in the previous section give for zero temperature: $\langle b^\dagger_{{\bf l}0}b_{{\bf l}0}\rangle=0.0931$. Thus, the influence of this term reduces to an effective decrease of the exchange constant approximately by 20 percent. Notice that the value of $\langle b^\dagger_{{\bf l}0}b_{{\bf l}0}\rangle$ is considerably grater than the bulk value $\langle b^\dagger_{{\bf l}l_x}b_{{\bf l}l_x}\rangle=0.0730$.

Now let us consider the influence of spin fluctuations in the transversal part of the interaction in Hamiltonian~(\ref{Hamiltonian3}),
\begin{equation}\label{Hti}
H_{ti}=J\sum_{\bf l}\left(S^-_{{\bf l}0}S^+_{{\bf l},-1}+S^+_{{\bf l}0}S^-_{{\bf l},-1}\right),
\end{equation}
where
$$S^+_{{\bf l},-1}=a^\dagger_{{\bf l},-1,\uparrow}a_{{\bf l},-1,\downarrow},\quad S^-_{{\bf l},-1}=a^\dagger_{{\bf l},-1,\downarrow}a_{{\bf l},-1,\uparrow}.$$
Using the perturbation theory we find
\begin{eqnarray}
\hat{G}'({\bf q}nl_xl'_x\sigma)&=&\hat{G}({\bf q}nl_xl'_x\sigma)
-\hat{G}({\bf q}nl_x,-1,\sigma)\nonumber\\
&&\quad\times\hat{\Sigma}({\bf q}n\sigma)
\hat{G}'({\bf q}n,-1,l'_x\sigma),\label{eqforG'}
\end{eqnarray}
where $\hat{G}'({\bf q}nl_xl'_x\sigma)$ is the Fourier transform of the Matsubara Green's function with the components
\begin{equation}\label{MGf}
G'_{mm'}({\bf q}\tau l_xl'_x\sigma)=-\langle{\cal T}a_{{\bf q}ml_x\sigma}(\tau)a^\dagger_{{\bf q}m'l'_x\sigma}\rangle.
\end{equation}
Here ${\cal T}$ is the chronological operator and the averaging and the dependence on the imaginary time $\tau$ are determined with the total Hamiltonian (\ref{Hamiltonian3}). In Eq.~(\ref{eqforG'}), the integer $n$ stands for the fermion Matsubara frequency $i\omega_n=(2n+1)\pi T$ and $\hat{G}({\bf q}nl_xl'_x\sigma)$ is the Green's function (\ref{GfG}) at this frequency. In the Born approximation the self-energy in Eq.~(\ref{eqforG'}) reads
\begin{eqnarray}
\Sigma_{mm_1}({\bf q}n\sigma)&=&J^2\frac{T}{N}\sum_{{\bf q}_1\nu}
\bigl[D_{\bar{m}\bar{m}_1}({\bf q}_1\nu 00)\nonumber\\
&+&(-1)^{m+m_1}D_{\bar{m}\bar{m}_1}({\bf q}_1'\nu 00)\bigr]\nonumber\\ &\times&G_{mm_1}({\bf q-q}_1,n-\nu,-1,-1,-\sigma),\quad \label{selfenergy}
\end{eqnarray}
where $\hat{D}({\bf q}\nu l_xl'_x)$ is the spin-wave Green's function (\ref{GfD}) at the boson Matsubara frequency $i\omega_\nu=2\nu\pi T$ and $\bar{m}=m\sigma+\frac{3}{2}(1-\sigma)$.

Notice that Eq.~(\ref{eqforG'}) has the same structure as Eq.~(\ref{eqforG2}). To estimate the value of the self-energy~(\ref{selfenergy}) we can use Eq.~(\ref{eqforG0}) for the electron Green's function and the approximation for the Green's function $\hat{D}({\bf q}\nu 00)$ which takes into account only the pole of the surface spin wave. Due to its dispersion and the integration over $k_x$ in Eq.~(\ref{eqforG0}) the Kondo-like divergencies are integrated out. As a result it can be seen that the self-energy~(\ref{selfenergy}) is of the order of $J^2/B$ and is much smaller than the respective multiplier $J$ in Eq.~(\ref{eqforG2}). Thus, the contribution of the interaction~(\ref{Hti}) can be neglected.

\section{Conclusion}
In this paper, we have considered elementary excitations of the heterostructure of the semi-infinite metal and the Mott insulator in the case when the crystals differ only in the value of the Hubbard repulsion -- it is zero in the metal, and it exceeds the critical value for the Mott transition in the insulator. At half-filling and low temperatures the insulator has the long-range antiferromagnetic order and its low-lying excitations are spin waves. We used the unitary transformation to reduce the initial Hamiltonian to a simpler one which describes the heterostructure of the metal and the Heisenberg antiferromagnet. At the interface of this heterostructure spins of the antiferromagnet interact with spins of electrons in the metal. We found that elementary excitations of the antiferromagnet are standing spin waves with the dispersion similar to that in the infinite case and the magnon mode localized near the surface of the antiferromagnet. This mode has the dispersion of the two-dimensional spin waves with somewhat increased for a given superexchange constant frequency. The description of this mode has much in common with the description of localized states near a point defect. Analogously to these states the mode ejects bulk modes -- the standing waves -- from the region near the surface. The antiferromagnetic order of the insulator induces the antiferromagnetic ordering in the metal. The magnetization in the metal decreases exponentially with distance from the interface. The correlation length is equal to 5--6 lattice spacings and depends only weakly on the parameters of the problem.

\begin{acknowledgments}
This work was partially supported by the ETF grant No.~6918.
\end{acknowledgments}

\end{document}